\documentclass[twocolumn,showpacs,preprintnumbers,amsmath,amssymb]{revtex4}
\usepackage{cases}
\usepackage{graphicx}
\usepackage{dcolumn}
\usepackage{bm}
\usepackage{amsmath}

\begin{document}

\title{ An effective method of calculating the non-Markovianity $\mathcal{N}$ for single channel open systems}
\author{Zhi He}
\author{Jian Zou}%
 \email{zoujian@bit.edu.cn}
 \author{Lin Li}
 \author{Bin Shao}%
 \affiliation{Department of
Physics, and Key Laboratory of Cluster Science of Ministry of
Education, Beijing Institute of Technology, Beijing 100081,
People's Republic of China }%
\date{\today}

\begin{abstract}
We propose an effective method which can simplify the optimization
of the increase of the trace distance over all pairs of initial
states in calculating the non-Markovianity $\mathcal{N}$ for
single channel open systems. For the amplitude damping channel, we
can unify the results of Breuer $et$ $al$. [Phys. Rev. Lett. \bf
103\rm, 210401 (2009)] in the large-detuning case and the results
of Xu $et$ $al$. [Phys. Rev. A \bf 81\rm, 044105 (2010)] in the
resonant case; furthermore, for the general off-resonant cases we
can obtain a very tight lower bound of $\mathcal{N}$.
 As another application of our method, we also discuss $\mathcal{N}$
 for the non-Markovian depolarizing channel.

\end{abstract}
\maketitle

\section{Introduction}
Inevitable interaction with external environment may lead to the
phenomenon of decoherence for open quantum systems. In general,
the assumption of Markovian approximation usually was applied to
dynamical evolution of system. However, recently people found
that the Markovian processes
without memory and non-Markovian processes with memory can lead to
distinctly different effects on decoherence and disentanglement of
open systems. Thus the non-Markovian dynamics have become
increasingly important and are extensive studied in both discrete
variable \cite{ Bellomo2007, Krovi2007, Piilo2008, Mazzola2009,
Zhang2009, Das2009, Sxu2010, Tan2010, Yeo2010} and
continuous-variable \cite{  Maniscalco2007, Liu2007, An2007,
Paz2008,  Shiokawa2009, Vasile2009} systems. It has been found
that the non-Markovian effect of environment can extend
significantly the entanglement time of the qubits
\cite{Bellomo2007} and has been experimentally observed \cite{
Sxu2010}. Because of the importance of the non-Markovian effect of
environment, some authors \cite{Wolf2008, Breuer2009, Rivas2010,
Ming912} have developed some measures to detect the
non-Markovianity of open systems from different points of view.
 Breuer $et$ $al$. \cite{Breuer2009} proposed
 a computable measure $\mathcal{N}$ to detect the non-Markovianity
 of open  systems. The idea is based on the distinguishability of
 quantum states which results from information flow between the open
system and its environment. Rivas $et$ $al$. \cite {Rivas2010}
also proposed a measure of non-Markovianity which is based on the
fact that negative rates are linked to whether entanglement
between the system and an ancilla can increase. Lu $et$ $al$.
\cite {Ming912} defined a measure of non-Markovianity using
quantum fisher information flow. As we know that evaluation of
$\mathcal{N}$ requires optimization of the total increase of the
trace distance over all pairs of initial states, which is very
difficult to accomplish. In Ref. \cite{Breuer2009} Breuer $et$
$al$. considered a two-level system interacting with a reservoir
which possesses the Lorentzian spectral property, and in the large
detuning case they found by numerical simulation the pair of
initial states $\rho_1(0)=|e\rangle\langle e|$ and
$\rho_2(0)=|g\rangle\langle g|$ which  make the optimization of
the total increase of the trace distance (It should be noted that
in Ref. \cite{Breuer2009} they use $|+\rangle$ and $|-\rangle$
instead of $|e\rangle$ and $|g\rangle$ to represent the excited
and ground states respectively). Later, for the same two initial
states Li $et$ $al$. \cite{Li2010} obtained the analytical
expression of the trace distance
$D(\rho_1(t),\rho_2(t))=|h(t)|^2$.
 Very recently for the same model in the resonant case, Xu $et$ $al$.
\cite{Xu2010} found two different initial states which make the
optimization of the total increase of the trace distance by using an
analytical method and the corresponding trace distance
$D(\rho_1(t),\rho_2(t))=|h(t)|$. It should be noted that in Ref.
\cite{Xu2010} they use $b(t)$ instead of $h(t)$ to represent the
amplitude damping of the excited state $|e\rangle$. It is intriguing
that these two results are quite different, that is, in these two
cases the two initial states which make the optimization are
different, and the corresponding trace distance are also different.
Because the optimization is very difficult to accomplish, until now
we have not seen any reports about the non-Markovianity for the
general off-resonant case. In Refs. \cite{Breuer2009, Xu2010} the
authors separately dealt with the optimization by different methods,
one is numerical and another is analytical. For the same model only
for different parameter regimes the results are quite different,
therefore an unified understanding of these results is in demands.

In this paper, we propose an effective method which can easily
optimize the increase of the trace distance over all pairs of
initial states in calculating the non-Markovianity $\mathcal{N}$
for single channel open systems. For the amplitude damping
channel, we analytically re-derive the results of Ref.
\cite{Breuer2009} in the large-detuning case and the results of
Ref. \cite{Xu2010} in the resonant case; furthermore, in the
general off-resonant cases we can obtain a very tight lower bound
for $\mathcal{N}$. Thus an unified understanding of the results of
Ref. \cite{Breuer2009} and Ref. \cite{Xu2010} is given. As another
application of our method, we also discuss $\mathcal{N}$  for the
non-Markovian depolarizing channel.

The paper is organized as follows. In Sec. II, we introduce our
method. The non-Markovian amplitude damping channel and the
non-Markovian depolarizing channel are examined by our method in
Sec. III. Finally, we give the conclusion of our results in
Sec. IV.

\section{ method of optimizing the total increase of trace distance }
Recently, Breuer $et$ $al$. \cite{Breuer2009} proposed a measure to detect the
non-Markovian behavior of quantum processes in open systems based on  the distinguishability of quantum states. The trace
distance $D$ describing the
distinguishability between the two states is defined as \cite{Nielsen2000}
\begin{equation}
D(\rho_1,\rho_2)=\frac{1}{2} \rm{tr}|\rho_1-\rho_2|,
\end{equation}
where $|M|=\sqrt{M^\dag M}$, and $0 \le D \le 1$. If $D=0$, the two
states are the same, and if $D=1$, the two states are totally
distinguishable.

Considering a quantum process $\Phi(t) $, $\rho(t)=
\Phi(t)\rho(0)$, where $\rho(0)$ and $\rho(t)$ denote the density
operators at time $t=0$ and at any time $t>0$, respectively, then
the non-Markovianity $\mathcal{N}$ is defined as
\begin{equation}
\mathcal{N}(\Phi) = \mathop {\rm{max}}\limits_{\rho_{1,2}(0)}
\int_{\sigma  > 0} {dt\sigma (t,\rho _{1,2} (0))},
\end{equation}
where $\sigma(t, \rho_{1, 2}(0))$ is the rate of change of the trace distance defined as
\begin{equation}
\sigma (t,\rho _{1,2} (0)) = \frac{d}{{dt}}D(\rho _1 (t),\rho _2 (t)).
\end{equation}
As we known that $\sigma (t,\rho _{1,2} (0))\leq 0$ corresponding
to all dynamical semigroups and all time-dependent Markovian
processes, and a process is non-Markovian if there exists a pair
of initial states and at certain time $t$ such that $\sigma
(t,\rho _{1,2} (0))>0$. Physically, this means that for
non-Markovian dynamics the distinguishability of the pair of
states increases at certain times.

In view of Eq. (3), the non-Markovianity $\mathcal{N}$ also can be written as the following form
\begin{equation}
\begin{array}{l}
 \mathcal{N}(\Phi) = \mathop {\max }\limits_{\rho _{1,2} (0)} \sum\limits_n {[D(\rho _1 (\tau _n^{\max } ),{\rm{ }}\rho _2 (\tau _n^{\max } ))}  \\
 ~~~~~~~~~~~~- D(\rho _1 (\tau _n^{\min } ),{\rm{ }}\rho _2 (\tau _n^{\min } ))],
 \end{array}
\end{equation}
where $\tau_n^{\rm{max}}$ and $\tau_n^{\rm{min}}$ correspond to the
time points of the local maximum and minimum of
$D(\rho_1(t),\rho_2(t))$, respectively. $\mathcal{N}(\Phi )$ can be
calculated as follows: One first derives the increment of the trace
distance over each time interval [$\tau_n^{\rm{min}}$,
$\tau_n^{\rm{max}}$] for any pairs of initial states, then sums up
the total contributions of all intervals, finally performs the
maximization for all pairs of initial states.

Generally speaking it is very difficult to make the maximization
in Eq. (4). In this paper we want to find easy ways to maximize
the increase of trace distance. Our idea is like this: First we
find the two specific initial states which make the maximization
of the quantity $\mathcal{N}_n(\Phi)$ at each time interval. The
$\mathcal{N}_n(\Phi)$ is defined as the difference between the
local maximum and local minimum of the trace distance for
arbitrary time interval [$\tau_n^{\rm{min}}$,
$\tau_n^{\rm{max}}$]. That is
\begin{equation}
\begin{array}{l}
 \mathcal{N}_n(\Phi) = \mathop {\max }\limits_{\rho _{1,2} (0)}{[D(\rho _1 (\tau _n^{\max } ),{\rm{ }}\rho _2 (\tau _n^{\max } ))}  \\
 ~~~~~~~~~~~~- D(\rho _1 (\tau _n^{\min } ),{\rm{ }}\rho _2 (\tau _n^{\min } ))].
 \end{array}
\end{equation}
Apparently it is much easier to find the two initial states which
makes the maximization in Eq. (5) than to find the two initial
states which makes the summation in Eq. (4) maximal. Then for
specific non-Markovian channel we try to prove that the two
initial states we found can also make the summation in Eq. (4)
maximal. Generally speaking it is not easy to prove this. If it
can not be proved, we still believe that the pair of initial
states which make the increase of the trace distance in single
time interval maximal, will also make the optimization of the
summation in $\mathcal{N}$. Of course this is not rigorous. If we
are strict enough, at least in this case we can find a lower bound
of $\mathcal{N}(\Phi)$, and we argue that this lower bound is
tight. We will show in the following that this method is very
effective.

\section{applications }
Based on the above idea, we can calculate the non-Markovianity
$\mathcal{N}$ for the non-Markovian amplitude damping channel and
the non-Markovian depolarizing channel.

\subsection{Non-Markovian amplitude damping channel }
We consider a two-level system (qubit) interacting with a zero temperature
reservoir.
The Hamiltonian of the total system under the rotating wave approximation is given by ($\hbar=1$)
\begin{equation}
\hat{H}=\omega_0\hat{\sigma}_+\hat{\sigma}_-+\sum_{k=1}^N\omega_k\hat{a}_k^{\dag}\hat{a}_k+\sum_{k=1}^N(g_k\hat{\sigma}_-\hat{a}_k^{\dag}+g_k^\ast\hat{\sigma}_+\hat{a}_k),
\end{equation}
where $\hat{\sigma}_+=|e\rangle\langle g|$ and  $\hat{\sigma}_-=|g\rangle\langle e|$, are the Pauli raising and lowering operators
 for the two-level system, respectively. $\omega_0$ is the Bohr frequency of the two-level system,
 $\hat{a}_k$ and  $\hat{a}_k^{\dag}$ are the annihilation and creation operators for reservoir mode $k$, $\omega_k$ is
 the frequency of the mode $k$ of the reservoir, and $g_k$ is the coupling constant. The Hamiltonian of Eq.(6) can
describe various systems.  For concrete discussion we take a two-level atom interacting with
the reservoir formed by the quantized modes of a high-$\mathcal{Q}$ cavity. The dynamics of the reduced density matrix
for the two-level atom can be written as  \cite{Breuer2002}
 \begin{equation}
\rho ^S (t) = \left( {\begin{array}{*{20}c}
   {\rho _{ee}^S (0)\left| {h(t)} \right|^2 } & {\rho _{eg}^S (0)h(t)}  \\
   {\rho _{eg}^{S * } (0)h^ *  (t)} & {1 - \rho _{ee}^S (0)\left| {h(t)} \right|^2 }  \\
\end{array}} \right)
\end{equation}
in the basis $\{|e\rangle,|g\rangle\}$, where the superscript S
represents the two-level atom. Corresponding $h(t)$ denotes the
amplitude of the upper level $|e\rangle$ of the atom initially
prepared in $|e\rangle$ and satisfies the following
integrodifferential equation
\begin{equation}
\frac{d}{{dt}}h(t) =  - \int_0^t {dt_1 f(t - } t_1 )h(t_1 ),
\end{equation}
where the kernel $f(t - t_1 ) = \int {d\omega J(\omega )} \exp
[i(\omega_0-\omega)(t - t_1 )]$ is related to the spectral density
$J(\omega)$ of the reservoir. The model describes the damping of a two-level atom in a cavity. In this
paper we restrict ourselves to the case that the atom-cavity
system has only one excitation, and suppose that $J(\omega)$ takes
the Lorentzian spectral density \cite{Breuer2002} with detuning,
namely
 \begin{equation}
J(\omega ) = \frac{1}{{2\pi }}\frac{{\gamma _0 \lambda ^2 }}{{(\omega_0-\delta  - \omega )^2  + \lambda ^2 }}.
\end{equation}
Here $\delta=\omega_0-\omega_c$ is the detuning of the center frequency of the cavity $\omega_c$ and
 the Bohr frequency of the two-level atom $\omega_0$,
the parameter $\lambda$ defines the spectral width of the coupling, which is associated with the
 reservoir correlation time by the relation $\tau_B=\lambda^{-1}$ and the parameter $\gamma_0$ is
 related to the relaxation time scale $\tau_R$ by the relation $\tau_R=\gamma_0^{-1}$. Therefore the
  analytic expression of $h(t)$ can be obtained as
\begin{equation}
h(t) = e^{ - (\lambda  - i\delta )t/2} [\cosh (dt/2) + (\lambda  - i\delta )\sinh (dt/2)/d]
\end{equation}
with $d=\sqrt{(\lambda-i\delta)^2-2\gamma_0\lambda}$.

 Based on the Hermiticity, and unit trace of a physical density matrix,
  any pair of initial states can be defined as \cite{Xu2010}
\begin{equation}
\begin{array}{l}
 \rho _1^S (0) = \left( {\begin{array}{*{20}c}
   \alpha  & \beta   \\
   {\beta ^ *  } & {1 - \alpha }  \\
\end{array}} \right) \\
 \rho _2^S (0) = \left( {\begin{array}{*{20}c}
   \mu  & \nu   \\
   {\nu ^ *  } & {1 - \mu }  \\
\end{array}} \right) \\
 \end{array}
\end{equation}
with $|\beta|^2 \le \alpha (1 - \alpha)$, $|\nu|^2 \le \mu(1-\mu)$
corresponding to the semipositivity of a density matrix, $\beta,\nu
\in \mathbb{C}$, $0 \le \alpha, \mu\le 1$, and $\alpha, \mu\in
\mathbb{R}$. Thus, the evolution of the corresponding density matrix
can be obtained
 \begin{equation}
\begin{array}{l}
 \rho _1^S (t) = \left( {\begin{array}{*{20}c}
   {\alpha \left| {h(t)} \right|^2 } & {\beta h(t)}  \\
   {\beta ^ *  h^ *  (t)} & {1 - \alpha \left| {h(t)} \right|^2 }  \\
\end{array}} \right) \\
 \rho _2^S (t) = \left( {\begin{array}{*{20}c}
   {\mu \left| {h(t)} \right|^2 } & {\nu h(t)}  \\
   {\nu ^ *  h^ *  (t)} & {1 - \mu \left| {h(t)} \right|^2 }  \\
\end{array}} \right). \\
 \end{array}
\end{equation}
The combination of Eqs. (1) and (12) immediately provides the expression of
the trace distance at any time $t\geq0$
\begin{equation}
D(\rho_1^S(t),\rho_2^S(t))=\sqrt{|h(t)|^4(\alpha-\mu)^2+|h(t)|^2|\beta-\nu|^2},
\end{equation}
which has been obtained in Ref. \cite{Xu2010}. It is noted that
the maximization of trace distance in the resonant case has been
given in Ref. \cite{Xu2010}, however, their method cannot be
extended to the general off-detuning case because the condition
$|h(\tau_n^{\min})|=0$ can not always be guaranteed at each local
minima. Now using Eq. (5), we can easily achieve the maximization
of $\mathcal{N}_n$ analytically,
  namely the optimization of the trace distance difference between the local maximum  and local minimum
in the time interval [$\tau_n^{\rm{min}}$, $\tau_n^{\rm{max}}$] by
choosing two specific initial states. According to Eqs. (5) and
(13), $\mathcal{N}_n$ can be written as
\begin{equation}
\begin{array}{l}
 \mathcal{N}_n = \mathop {\max }\limits_{\rho _{1,2} (0)}
 [ |h(\tau_n^{\rm{max}})|\sqrt{|h(\tau_n^{\rm{max}})|^2(\alpha-\mu)^2+|\beta-\nu|^2}  \\
 ~~~~~~~~~~~~- |h(\tau_n^{\rm{min}})|\sqrt{|h(\tau_n^{\rm{min}})|^2(\alpha-\mu)^2+|\beta-\nu|^2} ].
 \end{array}
\end{equation}
When $t=0$, $|h(0)|=1$, and from Eq. (13)
$D=\sqrt{(\alpha-\mu)^2+|\beta-\nu|^2}\leq1$. Furthermore, the
condition is equivalent to these parameterized conditions
$(\alpha-\mu)=r\rm{cos}\theta$,
$\beta-\nu=re^{i\phi}\rm{sin}\theta$ ($r\leq1$, $\theta\in[0, 2\pi
]$ and $\phi\in[0, \pi]$). Substituting these parameterized
conditions into Eq. (14) and considering that the maximization is
over all pairs of initial states, we can obtain that the
maximization condition requires
$\sqrt{(\alpha-\mu)^2+|\beta-\nu|^2}=1$ corresponding to $r=1$.
Then the problem becomes the maximization of the following
$\mathcal{N}_n^o(\theta)$
\begin{equation}
\begin{array}{l}
 \mathcal{N}_n^o(\theta) = |h(\tau_n^{\rm{max}})|\sqrt{|h(\tau_n^{\rm{max}})|^2\rm{cos}^2\theta+\rm{sin}^2\theta}  \\
 ~~~~~~~~~~~~- |h(\tau_n^{\rm{min}})|\sqrt{|h(\tau_n^{\rm{min}})|^2\rm{cos}^2\theta+\rm{sin}^2\theta}.
 \end{array}
\end{equation}
From the following equation
\begin{equation}
\frac{{\partial N_n^o(\theta)}}{{\partial \theta }} = 0
\end{equation}
we can obtain the extrema, which are $\mathcal{N}_{n1}^o=A^2-B^2$
when $\theta=0$; $\mathcal{N}_{n2}^o=A-B$ when $\theta=\pi/2$ or
$\theta=3\pi/2$;
$\mathcal{N}_{n3}^o=A^2\sqrt{\frac{1-A^2}{(B^2-1)(A^2+B^2-1)}}-B^2\sqrt{\frac{1-B^2}{(A^2-1)(A^2+B^2-1)}}$
when $\theta=\rm{arccos} [ \sqrt{\frac
{-A^2+2A^4-A^6+B^2-2B^4+B^6}{(A^2-1)(B^2-1)(A^4-A^2+B^2-B^4)}}] $,
where $A=|h(\tau_n^{\rm{max}})|$, $B=|h(\tau_n^{\rm{min}})|$ and
$A > B$. So $\mathcal{N}_n$ can be represented as
\begin{equation}
\mathcal{N}_n={\rm{max}}\{\mathcal{N}_{n1}^o, \mathcal{N}_{n2}^o, \mathcal{N}_{n3}^o\}.
\end{equation}
From numerical calculation, we find that for any $A$ and $B$
satisfying $0 < B< A < 1$, $\mathcal{N}_{n3}^o$ is always less
than $\mathcal{N}_{n1}^o$ and $\mathcal{N}_{n2}^o$, so
$\mathcal{N}_n={\rm{max}}\{\mathcal{N}_{n1}^o,
\mathcal{N}_{n2}^o\}$. From the definitions of $\mathcal{N}_{n1}^o$
and $\mathcal{N}_{n2}^o$  we can obtain that when $0\leq A+B<1$,
$\mathcal{N}_{n1}^o<\mathcal{N}_{n2}^o$; when $A+B=1$,
$\mathcal{N}_{n1}^o$ = $\mathcal{N}_{n2}^o$; when $A+B>1$,
$\mathcal{N}_{n1}^o>\mathcal{N}_{n2}^o$.

(i) In the resonant case, that is $\delta=0$, it is obvious that
$A+B=|h(\tau_n^{\rm{max}})|+|h(\tau_n^{\rm{min}})|<1$ because
$B=|h(\tau_n^{\rm{min}})|=0$ at $\tau_n^{\rm{min}} =
2[n\pi-\rm{arctan}(d't/2)]/d'$ with
  $n$=1,2,3,..., and $d'=\sqrt{|\lambda^2-2\gamma_0\lambda|}$.
  So in any time interval [$\tau_n^{\rm{min}}$,
$\tau_n^{\rm{max}}$], $\mathcal{N}_n={\rm{max}} \{
\mathcal{N}_{n1}^o, \mathcal{N}_{n2}^o \}
=\mathcal{N}_{n2}^o=A-B=|h(\tau_n^{\rm{max}})|$. For
$\mathcal{N}_{n2}^o$, $\theta=\pi/2$ or $3\pi/2$, and the two
initial states correspond to $\alpha=\mu$ and $|\beta-\nu|=1$.
Because in this case $\mathcal{N}_n$ reaches its maximum for any
$n$, the same two initial states, that is $\alpha=\mu$ and
$|\beta-\nu|=1$, are also the two initial states which make the summation in Eq. (4) maximal.
 It is worth noting that these conditions
$\alpha=\mu, |\beta-\nu|=1$
 together with $|\beta|^2 \le \alpha (1 - \alpha)$ and $|\nu|^2 \le \mu(1-\mu)$ are equivalent to
 the conditions $\alpha=\mu=1/2$, $|\beta|=|\nu|=1/2$
 and $|\beta-\nu|=1$ obtained in Ref. \cite{Xu2010}, which can be explained as follows. Our conditions
 can be changed into $\alpha=\mu$, $|\beta-\nu|=1$, $(\alpha-1/2)^2+|\beta|^2\leq(1/2)^2$
 and $(\mu-1/2)^2+|\nu|^2\leq(1/2)^2$. From the geometric point of view, the new conditions
 indicate that the two points ($\alpha, |\beta|$)
 and ($\mu, |\nu|$) is in (or on the circumference of ) the same circle
\begin{equation}
(x-1/2)^2+|y|^2=(1/2)^2
\end{equation}
with $x\in\mathbb{R}$ and $y\in\mathbb{C}$. It is easy to check
that the conditions $\alpha=\mu$, and $|\beta-\nu|=1$ are just
$\alpha=\mu=1/2$, $|\beta|=|\nu|=1/2$, $|\beta-\nu|=1$. In
summary, in the resonant case our results are consistent with the
results of Ref. \cite{Xu2010}.

(ii) In the off-resonant case, that is $\delta\neq0$,
$A+B=|h(\tau_n^{\rm{max}})|+|h(\tau_n^{\rm{min}})|$ may be less
than 1, equal to 1, or more than 1 depending on the values of
$\gamma_0, \lambda $ and $\delta$. We have proved that depending
on the value of $A+B$ there are only two pairs of initial states
which make the maximization of $\mathcal{N}_n$ for each $n$. Next
we give our effective and practical method to calculate
$\mathcal{N}$ for any fixed parameters $\gamma_0$, $\lambda $ and
$\delta$. From Eqs. (2) and (3) we can use the two pairs of
initial states to obtain $\mathcal{N}_1$ and $\mathcal{N}_2$
respectively, $\mathcal{N}_1=\int_{\sigma_1>0} dt \sigma_1( t,
\theta = 0)$, $\mathcal{N}_2=\int_{\sigma_2>0} dt \sigma_2(t,
\theta = \pi/2$ or $3\pi/2)$. Correspondingly, the expressions of
$\sigma_1( t, \theta = 0)$ and $\sigma_2(t, \theta = \pi/2$ or
$3\pi/2)$ are given by

 \begin{equation}
\sigma_1(t)=e^{-\lambda t}\{\mu[\cosh(at)-\cos(bt)]+\nu \sinh(at)-\xi\sin(bt)\},
\end{equation}
\begin{equation}
\sigma_2(t)=\frac{e^{-\frac{\lambda t}{2}}\{\mu[\cosh(at)-\cos(bt)]+\nu \sinh(at)-\xi\sin(bt)\}}
{2\sqrt{{ \eta\sinh(at)+\chi\sin(bt)+\kappa\cosh(at)+\varsigma\cos(bt)}}},
\end{equation}
where $a$ and $b$ denote the real part and imaginary part of $d$ respectively;
$\mu=\frac{1}{2|d|^2}(\lambda a^2-\lambda b^2-\lambda \delta^2-\lambda^3-2ab\delta)$,
 $\nu=\frac{1}{2|d|^2}(a^3+ab^2-\lambda^2a+a\delta^2+2b\delta\lambda)$,
 $\xi=\frac{1}{2|d|^2}(b^3+ba^2+\lambda^2b-b\delta^2+2a\delta\lambda)$,
 $\eta=\frac{1}{2|d|^2}(2a\lambda-2b\delta)$,  $\chi=\frac{1}{2|d|^2}(2b\lambda+2a\delta)$,
 $\kappa=\frac{1}{2|d|^2}(\lambda^2+\delta^2+a^2+b^2)$, $\varsigma=\frac{1}{2|d|^2}(a^2+b^2-\lambda^2-\delta^2)$;
 $|d|$ denotes the absolute value of $d$. It is worth noting that Eq. (19) has been obtained in Ref. \cite{Li2010}.
We can not prove but we believe that one of the two pairs of
initial states we found can also make the optimization in the
summation of Eq. (4), thus $\mathcal{N}=\max
\{\mathcal{N}_1,\mathcal{N}_2\}$. If we are strict enough, at
least it is a very tight lower bound (TLD) for the genuine
$\mathcal{N}$,
 \begin{equation}
\mathcal{N}_{\rm{TLD}}={\rm{max}}\{\mathcal{N}_1, \mathcal{N}_2 \}.
\end{equation}

We plot $\mathcal{N}_{\rm{TLD}}$, $\mathcal{N}_1$,
 and $\mathcal{N}_2$ as functions of
detuning $\delta$ for $\lambda=0.1\gamma_0$ in Fig. 1, and
 $\mathcal{N}_{\rm{TLD}}$, $\mathcal{N}_1$,
 and $\mathcal{N}_2$ as functions of $\lambda$ for $\delta=0.1\gamma_0$ in Fig. 2.
 From Fig. 1 we can see that
there exists a critical point for $\delta$ at which the pair of
initial states change from $\theta=\pi/2$ or $\theta=3\pi/2$ to
$\theta=0$. More specifically
 when $\delta<\delta_c$, $\mathcal{N}_{\rm{TLD}}=\mathcal{N}_2 $
corresponding to the two initial states $\theta=\pi/2$ or
$\theta=3\pi/2$; when $\delta=\delta_c$, $
\mathcal{N}_{\rm{TLD}}=\mathcal{N}_1=\mathcal{N}_2$ corresponding
to the two initial states $\theta=0$, and $\theta=\pi/2$ or
$3\pi/2$; when $\delta>\delta_c$,
$\mathcal{N}_{\rm{TLD}}=\mathcal{N}_1$ corresponding to the two
initial states $\theta=0$. Similarly, it can be seen from Fig. 2
that there also exists a critical point for $\lambda$ at which the
pair of initial states change from $\theta=0$ to $\theta=\pi/2$ or
$\theta=3\pi/2$.

\begin{figure}
\centering
\includegraphics[width=7.5cm]{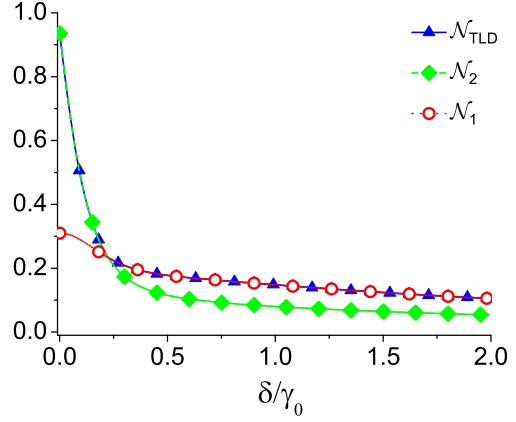}
\hspace{0.5cm} \caption{ (Color online) $\mathcal{N}_{\rm{TLD}}$,
$\mathcal{N}_1$ and $\mathcal{N}_2$ as a function of $\delta$,
$\lambda=0.1\gamma_0$. }
\end{figure}
\begin{figure}
\centering
\includegraphics[width=7.5cm]{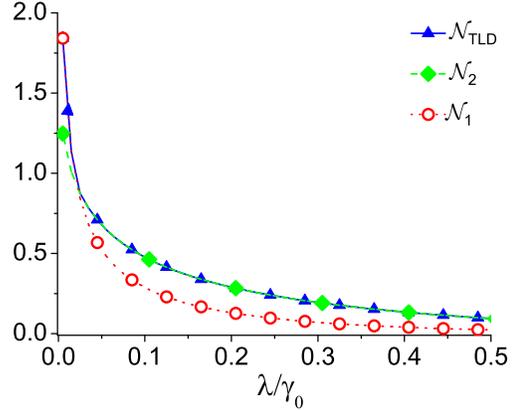}
\hspace{0.5cm} \caption{(Color online) $\mathcal{N}_{\rm{TLD}}$,
$\mathcal{N}_1$ and $\mathcal{N}_2$ as a function of $\lambda$,
$\delta=0.1\gamma_0$. }
\end{figure}

It is worth noting that in the large detuning case by numerical
simulation Breuer $et$ $al$. \cite{Breuer2009} have performed the
optimization of the total increase of the trace distance. In the
large detuning case from Eq. (10) we know
$A+B=|h(\tau_n^{\rm{max}})|+|h(\tau_n^{\rm{min}})|>1$ for any
$n=1,2,3,...$. Therefore, we can obtain $\mathcal{N}_n
={\rm{max}}\{\mathcal{N}_{n1}^o,
\mathcal{N}_{n2}^o\}=\mathcal{N}_{n1}^o=A^2-B^2=
|h(\tau_n^{\rm{max}})|^2-|h(\tau_n^{\rm{min}})|^2$ for any $n$
  and the two initial states satisfy
   $\alpha-\mu=1$ and $|\beta-\nu|=0$ corresponding to $\theta=0$.
Because the two initial states ($\theta =0$) make the optimization
of $\mathcal{N}_n$ for any $n$, they also make the optimization of
the summation in Eq. (4).
 Similar to the resonant case,
the optimization conditions including $\alpha-\mu=1$,
$|\beta-\nu|=0$, $|\beta|^2 \le \alpha (1 - \alpha)$, and $|\nu|^2
\le \mu(1-\mu)$ can be simplified into $\alpha=1$, $\mu=0$ and
$\beta=\nu=0$.  Evidently, the two initial states are
$\rho_1(0)=|e\rangle\langle e|$ and $\rho_2(0)=|g\rangle\langle
g|$ which has been obtained in Ref. \cite{Breuer2009}. It is noted
that the condition is obtained by numerical simulation in
\cite{Breuer2009}, but here we obtain the same condition using an
analytic method. Then we can also obtain the trace distance
$D=|h(t)|^2$ for these two initial states which is given by Ref.
\cite{Li2010}. In summary, in the large-detuning case our results
are consistent with the results reported in Refs.
\cite{Breuer2009, Li2010}.

\subsection{Non-Markovian depolarizing channel }
As another application of our method, we consider the
non-Markovianity $\mathcal{N}$ for a non-Markovian depolarizing
channel. The dynamical property of this system and in particular
the conditions of complete positivity of the map corresponding to
a master equation  have been studied in detail by Daffer $et$
$al$. in Ref. \cite{Daffer2004}. For this model the time-dependent
Hamiltonian, that corresponds to a two-level system subjected  to
  random telegraphic noise, is
 \begin{equation}
H(t)=\hbar\sum_{i=1}^3\Gamma_i(t)\sigma_i,
\end{equation}
where $\Gamma_i(t)=a_in_i(t)$ are independent random variables,
and $\sigma_i$ are the usual Pauli operators. $n_i(t)$ has a
Poisson distribution with  a mean equal to $t/2\tau_i$, while
$a_i$ is an independent coin-flip random variable taking the
values $\pm a_i$. For the time-dependent Hamiltonian of Eq. (22),
the corresponding equation of motion for the density operator is
govern by the von Neumann equation
$\dot{\rho}=-i/\hbar[H(t),\rho]=-i\Sigma_k\Gamma_k(t)[\sigma_k,\rho]$
which has the following formal solution
\begin{equation}
\rho(t)=\rho(0)-i\int_0^t\sum_k\Gamma_k(s)[\sigma_k,\rho(s)]ds.
\end{equation}
Substituting the formal solution Eq. (23) into the von Neumann equation and performing a stochastic average, one can obtain
the following memory kernel master equation

\begin{equation}
\dot{\rho}(t)=-\int_0^t\sum_ke^{-(t-t')/\tau_k}a_k^2[\sigma_k,[\sigma_k,\rho(t')]]dt',
\end{equation}
where the correlation function of the random telegraph signal
$\langle\Gamma_j(t)\Gamma_k(t')\rangle=a_k^2e^{-|t-t'|/\tau_k}\delta_{jk}$
contributes to the memory kernel.  It has been pointed out
\cite{Daffer2004} that the system density operator with an
exponential memory kernel obeys a homogeneous Volterra equation
after averaging over the reservoir variables, and also proved that
when two of $a_i$ are zero, only one direction having the noise,
the map $\Phi(\rho)$ can be written as Kraus operator form \cite
{Kraus1983}, namely
\begin{equation}
\rho'(t)=\Phi_t ( \rho)= \sum_{k=1}^4 A_k^ \dag \rho A_k.
\end{equation}
For simplicity, in this paper we only consider the case the $z$
direction with noise, $x$ and $y$ directions without noise.
Therefore, $A_1=0$, $A_2=0$,
$A_3=\sqrt{[1-\Lambda(\nu)]/2}\sigma_3$, and
$A_4=\sqrt{[1+\Lambda(\nu)]/2}I$, where $\Lambda(\nu)
=\rm{exp}(-\nu)[\rm{cos}(\mu\nu)+\rm{sin}(\mu\nu)/\mu]$ with
$\mu=\sqrt{(4a\tau)^2-1}$, and $\nu=t/2\tau$ is a dimensionless
time. $a$ is the coupling strength of the system with the external
environment while $\tau$ determines which frequencies the system
prefers most. For convenience we let $\lambda=1/\tau$, then
$\Lambda(\nu)$ can be rewritten as
\begin{equation}
\begin{array}{l}
 \Lambda (t) =  \\
 \left\{ {\begin{array}{*{20}c}
   {e^{ - \lambda t/2} [\cosh (\frac{{\varepsilon t}}{2}) + \frac{\lambda }{\varepsilon }\sinh (\frac{{\varepsilon t}}{2})]~~~(16a^2  < \lambda^2 {\rm{) }}}  \\
   \\
   {e^{ - \lambda t/2} [1 + \frac{{\lambda t}}{2}]~~~~~~~~~~~~~~~~~~~~~(16a^2  = \lambda^2 {\rm{)}}}  \\
   \\ {e^{ - \lambda t/2} [\cos (\frac{{\varepsilon t}}{2}) + \frac{\lambda }{\varepsilon }\sin (\frac{{\varepsilon t}}{2})]~~~~~~(16a^2  > \lambda^2 {\rm{) }}},
\end{array}} \right.
 \end{array}
\end{equation}
where $\varepsilon=\sqrt{|16a^2-\lambda^2|}$.

Using the same two initial states of Eq. (11), from Eq. (25) we can
obtain the evolutions of the two density matrixes, respectively
\begin{equation}
\begin{array}{l}
 \rho _1^S (t) = \left( {\begin{array}{*{20}c}
   \alpha  & {\beta \Lambda(t) }  \\
   {\beta ^ *  \Lambda^*(t)} & {1 - \alpha }  \\
\end{array}} \right) \\
 \rho _2^S (t) = \left( {\begin{array}{*{20}c}
   \mu  & {\nu \Lambda(t) }  \\
   {\nu ^ *  \Lambda^*(t)} & {1 - \mu }  \\
\end{array}} \right). \\
 \end{array}
\end{equation}
Therefore the trace distance can be obtained
 \begin{equation}
D(\rho_1^S(t),\rho_2^S(t))=\sqrt{(\alpha-\mu)^2+|\Lambda(t)|^2|\beta-\nu|^2}.
\end{equation}
From Eqs. (5) and (28), $\mathcal{N}_n$ can be expressed as
\begin{equation}
\begin{array}{l}
 \mathcal{N}_n = \mathop {\max }\limits_{\rho _{1,2} (0)}[ \sqrt{(\alpha-\mu)^2+|\Lambda(\tau_n^{\rm{max}})|^2|\beta-\nu|^2}  \\
 ~~~~~~~~~~~~- \sqrt{(\alpha-\mu)^2+|\Lambda(\tau_n^{\rm{min}})|^2|\beta-\nu|^2} ].
 \end{array}
\end{equation}
As in the subsection A, after parameterizing $(\alpha-\mu)$ and
$|\beta-\nu|$, the problem becomes the maximization of the following
$\mathcal{N}_n^o(\theta)$
\begin{equation}
\begin{array}{l}
 \mathcal{N}_n^o(\theta) = \sqrt{\rm{cos}^2\theta+|\Lambda(\tau_n^{\rm{max}})|^2\rm{sin}^2\theta}  \\
 ~~~~~~~~~~~~- \sqrt{\rm{cos}^2\theta+|\Lambda(\tau_n^{\rm{min}})|^2\rm{sin}^2\theta}.
 \end{array}
\end{equation}
Then we can obtain the extrema of $\mathcal{N}_n^o$:
 $\mathcal{N}_{n1}^o=0$ when $\theta=0$; $\mathcal{N}_{n2}^o=A-B$ when $\theta=\pi/2$ or $3\pi/2$,
 where $A=|\Lambda(\tau_n^{\rm{max}})|$, $B=|\Lambda(\tau_n^{\rm{min}})|$ and $A>B$.
Obviously, $\mathcal{N}_n={\rm{max}}\{\mathcal{N}_{n1}^o,
\mathcal{N}_{n2}^o\}=\mathcal{N}_{n2}^o=A-B $. Because the pair of initial states
 corresponding to $\theta=\pi/2$ or $3\pi/2$ makes the increase of trace
 distance $\mathcal{N}_n^o$ maximal for any $n$, it also makes the summation in Eq. (4) maximal. The condition $\theta=\pi/2$ or
$3\pi/2$ means $\alpha-\mu=0$ and $|\beta-\nu|=1$, and similar to
the discussion in subsection A the condition
 together with $|\beta|^2 \le \alpha (1 - \alpha)$ and $|\nu|^2 \le \mu(1-\mu)$ is also equivalent
to $\alpha=\mu=1/2$, $|\beta|=|\nu|=1/2$, $|\beta-\nu|=1$. The
trace distance after choosing the two initial states can be given
as
\begin{equation}
D=|\Lambda(t)|.
\end{equation}
It is noted that there are some similarity between the
non-Markovian depolarizing channel and the amplitude damping
channel. In the non-Markovian
depolarizing channel the maximal trace distance is a function of the
decoherence factor $|\Lambda(t)|$, while in the amplitude damping channel it is
a function of the amplitude damping factor $|h(t)|$.
 It is very interesting that the trace distance of the non-Markovian depolarizing channel
 for the specific pair of initial states which make the optimaization,
 is very similar to that
 of the amplitude damping channel in the resonant case, that is for the former the trace distance for the specific pair of initial states is equal to
 the decoherence factor $|\Lambda(t)|$, while for the latter it is equal to the amplitude damping factor $|h(t)|$;
 furthermore, in both cases the two initial states which makes the optimization are the same.
\begin{figure}
\centering
\includegraphics[width=6.0cm]{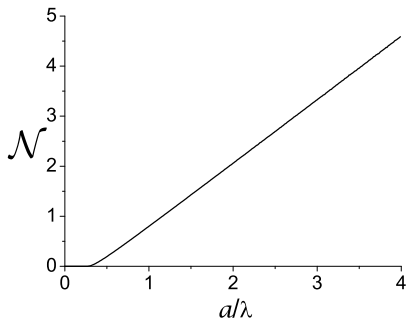}
\hspace{0.5cm}
\caption{The non-Markovianity $\mathcal{N}$ as a function of $a$. }
\end{figure}

We can obtain the rate of the change of the trace distance using Eq.
(3)
\begin{equation}
\sigma (t,\rho _{1,2} (0))=\frac{\Lambda(t)\frac{d}{{dt}}\Lambda(t)}{{|\Lambda(t)|}}.
\end{equation}
Having these preparations, we can easily calculate the
non-Markovianity $\mathcal{N}$ from Eq. (2). We plot
$\mathcal{N}$ as a function of $a$ in Fig. 3. From Fig. 3 we can
see that there is a threshold $a_c=\lambda/4$. When $a\leq a_c$,
$\mathcal{N}=0$, which means that the process is Markovian. While
when $a>a_c$, $\mathcal{N}$ increases with $a$, which means that
the process is non-Markovian. This can be easily understood:
 because $a$ represents the coupling strength of the system and the reservoir,
 it is obvious that the non-Markovianity $\mathcal{N}$
  will become more and more strong with  the increasing of $a$ in the non-Markovian regime.
It is worth noting that recently Mazzola \it et al. \rm \cite
{Mazzola2010} elucidated that the memory kernel master equation
does not ensure the presence of non-Markovian behavior in the
evolution of dynamics. For example, in Ref. \rm \cite
{Mazzola2010} they have shown that the non-Markovian behavior does
not appear in the memory kernel master equations Eqs.(4) and (10)
of the same reference. $\xi_{M}(R,t)$ and $\xi_{P}(R,t)$ are two functions corresponding to
two models Eq. (4) and Eq. (10) of Ref. \rm \cite {Mazzola2010}
respectively, which play the central role in the dynamics of the
system. It has been proved that $\xi_{M}(R,t)$ and $\xi_{P}(R,t)$
are positive,
 monotonically decreasing functions under the condition of positivity of the dynamical maps \rm \cite
 {Maniscalco2007-2}. Thus the quantum processes corresponding to Eqs.(4) and
 (10) of Ref. \rm \cite {Mazzola2010} are Markovian. However, the model we used can also be
described by the memory kernel master equation Eq. (24), but
 we clearly see that the function $\Lambda(t)$ which plays a central role in our model, isn't a monotonous function with
 time but is a damped oscillating function in some parameter regimes, which can be seen from Eq.
 (26). And in this case the quantum process we consider is non-Markovian.

\section{conclusions}
As we know that the definition of the non-Markovianity
$\mathcal{N}$ needs an optimization over all pairs of initial
states, and generally it is very hard to do this. In this paper we
proposed a method which can simplify this optimization. The main
idea is like this: First we find the pair of initial states which
make the maximization of the difference between the local maximum
and local minimum of the trace distance in arbitrary $n$th time
interval, then we try to prove that this pair of initial states
can also make the optimization of the summation over all pairs of
initial states in calculating the non-Markovianity $\mathcal{N}$.
Using this method we have analytically obtained the pair of
initial states which make the optimization in Eq. (4) and the
corresponding trace distance for the amplitude damping channel in
both the resonant case and the large-detuning case, and unified
the results of Breuer $et$ $al$. \cite{Breuer2009} and Xu $et$
$al$ \cite{Xu2010}; and we also have analytically obtained the
pair of initial states which make the optimization in Eq. (4) and
the corresponding trace distance for the non-Markovian
depolarizing channel. Generally speaking it can not always be
proved that the pair of initial states which make the maximization
of the $n$th difference between the local maximum and local
minimum of the trace distance, also make the optimization of the
summation in $\mathcal{N}$. But here we argue that the pair of
initial states which make the increase of the trace distance
maximal in one time interval, will also makes the optimization of
the summation in $\mathcal{N}$. We can not prove this in this
paper, and this needs further investigations. If we are strict
enough, at least we can obtain a very tight lower bound for
$\mathcal{N}$. For example, for the amplitude damping channel, we
have found only two pairs of initial states depending on the
system parameters which make the increase of the trace distance
maximal in arbitrary single time interval. Then it is easy to
calculate $\mathcal{N}$ for these two pairs of initial states, and
the larger one is the tight lower bound of $\mathcal{N}$. In this
paper we mainly focus on the single channel case and the
generalization to the general multi-channel case may be a more
challenging task. In one word we have simplified the problem of
finding a pair of initial states which optimizing the summation of
the increase of trace distance in many time interval into a
problem of optimizing the increase of trace distance in just one
single time interval.

\begin{acknowledgments}
This work is financially supported by National Science Foundation
of China (Grants No. 10974016, 11005008, and 11075013).
\end{acknowledgments}

\end{document}